\documentstyle[12pt]{article}

\makeatletter \@addtoreset{equation}{section} \makeatother

cm \oddsidemargin .5 true cm \evensidemargin .5 true cm

\def\Fock{|0\rangle\langle 0|}
\def\al{\alpha}
\def\ab{\bar{\alpha}}
\def\bb{\bar{\beta}}

\def\dab{\dot{\bar{\alpha}}}
\def\dbb{\dot{\bar{\beta}}}
\def\dbg{\dot{\bar{\gamma}}}
\def\dg{\bar{\gamma}}

\title{Nt-yet}

\makeatletter \@addtoreset{equation}{section} \makeatother

\newcommand{\be}{\begin{equation}}
\newcommand{\ee}{\end{equation}}
\newcommand{\bee}{\begin{eqnarray}}
\newcommand{\beee}{\begin{array}}
\newcommand{\eee}{\end{eqnarray}}
\newcommand{\eeee}{\end{array}}

\newcommand{\ga}{\alpha}
\newcommand{\gb}{\beta}
\newcommand{\gga}{\gamma}

\newcommand{\M}{{\cal M}}

\newcommand{\gd}{\delta}
\newcommand{\gl}{\lambda}

\newcommand{\gs}{\sigma}

\newcommand{\nn}{\nonumber}

\newcommand{\p}{\partial}
\renewcommand{\L}{{L}}

\newcommand{\f}{\frac}

\begin{document}

\begin{flushright}
\vspace{1mm} FIAN/TD/1-03\\
 January {2003}\\
\end{flushright}

\vspace{1cm}

\begin{center}
{\large\bf Free Field Dynamics in the Generalized $AdS$
(Super)Space} \vglue 0.6  true cm \vskip1cm {\bf
V.E.~DIDENKO\footnote{E-mail: didenko@lpi.ru} and
M.A.~VASILIEV\footnote{E-mail: vasiliev@lpi.ru}} \vglue 0.3  true
cm

I.E.Tamm Department of Theoretical Physics, Lebedev Physical Institute,\\
Leninsky prospect 53, 119991, Moscow, Russia \vskip2cm
\end{center}

\begin{abstract}
\baselineskip .4 true cm \noindent Pure gauge representation for
general vacuum background fields (Cartan forms) in the generalized
$AdS$ superspace identified with $OSp(L,M)$ is found. This allows
us to formulate dynamics of free massless fields in the
generalized $AdS$ space-time and to find their (generalized)
conformal and higher spin field transformation laws. Generic
solution of the field equations is also constructed explicitly.
The results are obtained with the aid of the star product
realization of  ortosymplectic superalgebras.
\end{abstract}

\section {Introduction}

In the recent papers \cite{osp, Sp} it was shown that infinite
multiplets of massless higher spins in $4d$ flat Minkowski
space-time admit description in terms of ten-dimensional
space-time $\M_4$ with real symmetric bispinor matrix coordinates
$X^{\ga\gb}=X^{\gb\ga}$, ($\al, \beta=1\dots4$). A single scalar
field $c(X)$ in $\M_4$ describes massless fields of all integer
spins in $4d$ Minkowski space-time upon imposing  field equations
found in \cite{osp}. Half-integer spin massless fields are
described analogously by a spinor field $c_\ga (X)$. That massless
fields of all spins have to admit some formulation in $\M_4$ was
argued by Fronsdal in the pioneering paper \cite{Fr} where it was
also stressed that such infinite sets of massless fields have to
form representations of the extension of the $4d$ conformal group
$su(2,2)$ to $sp(8|R)$. Then in \cite{BLS} it was found that
world-line particle models based on $sp(8)$ give rise to  massless
higher spin excitations of all spins. The explicit realization of
the $sp(8)$ symmetry by local transformations  was given in
\cite{osp} as well as the generalization of the proposed $sp(8)$
covariant dynamical equations to $\M_M$  with arbitrary even  $M$.

Properties of the $Sp(2M)$ invariant space-time $\M_M$ were
analyzed in \cite{Sp}. It was shown that the classical solutions
of the field equations define a causal structure and give rise to
correct quantization in a positive definite Hilbert space. Usual
$d$- dimensional Minkowski space-time appears as a  subspace of
the generalized space-time. The analysis of \cite{osp,Sp} was
performed for the case of flat space-time although the formalism
as a whole works in any (generalized) conformally flat background.
In particular it is interesting to extend this analysis to the
generalized anti-de Sitter space-time which was argued in
\cite{osp} to be the group manifold $Sp(M)$ ($M$ is even) having
$Sp(M) \times Sp(M) \subset Sp(2M)$ as the group of motions
realized by left and right group actions on itself. Since the
analysis of $Sp(2M)$ invariant higher spin systems is most
naturally performed in terms of star product algebras, for its
extension to the generalized $AdS$ space-time it is necessary to
built star-product realizations of left invariant Cartan forms
(i.e., flat connections) on $Sp(M)$. This is the primary goal of
this paper. Obtained results will allow us to present explicit
formulae for symmetries and solutions of the massless field
equations in the generalized $AdS$ space-time. The analogous
construction will also be given for the supersymmetric case
associated with $OSp(L,M)$.

Let us note that since the star product formalism we apply leads
to compact expressions for $OSp(L,M)$ Cartan superforms, apart
from the higher spin problem, the results obtained in this paper
may have applications to other problems where left-invariant forms
of $OSp(L,M)$  appear. For example, in \cite{tw} it was shown how
$OSp(1,M)$ Cartan forms can be used to construct twistor-like
actions for superparticles and possible applications to
superbranes were discussed while in \cite{Bars} a toy model of $M$
theory based on $osp(1,64)$ was suggested.

\subsection {Generalized conformal symmetry}

The generators $L_{mn}, P_{m}, K_{m}, D$ of the conformal algebra
$o(d,2)$  satisfy the following commutation relations
\bee
[L_{ab},L_{cd}]&=&\eta_{ac}L_{bd}-\eta_{bc}L_{ad}
+\eta_{ad}L_{cb}-\eta_{bd}L_{ca}\,,\nn \\
{[}L_{ab},P_{c}]=\eta_{ac}P_{b}-\eta_{bc}P_{a}\,, &\quad&
[L_{ab},K_{c}]=\eta_{ac}K_{b}-\eta_{bc}K_{a}\,,\nn\\
{[}L_{ab},D]=[P_{a},P_{b}]&=&[K_{a},K_{b}]=0\,,\nn\\
{[}P_{a},K_{b}]=2(\eta_{ab}D+L_{ab})\,, &\quad& [P_{a},D]=P_{a}\,,
\quad [K_{a},D]=-K_{a}\,,
\eee
$m,n = 0,\ldots d-1$, $\eta_{mn}=diag(1,-1\ldots -1) $. The
conformal  algebra can be realized by the vector fields

\bee
L_{ab}&=&\eta_{ac}x^{c}\frac{\partial}{\partial
x^{b}}-\eta_{bc}x^{c}\frac{\partial}{\partial x^{a}}\,,\nn \\
{P}_{a}&=&\frac{\partial}{\partial x^{a}},
\quad D=x^{a}\frac{\partial}{\partial x^{a}}\,,\nn \\
{K}_{a}&=&2\eta_{ac}x^{c}x^{b}\frac{\partial}{\partial x^{b}}
-\eta_{bc}x^{b}x^{c}\frac{\partial}{\partial x^{a}}\,.
\eee

The Poincare subalgebra is spanned by $L_{mn}$ and $P_{m}$.
$K_{m}$ and $D$ are the generators of special conformal
transformations and dilatations, respectively. To embed the
$AdS_{d}$ algebra $o(d-1,2)$ into the $d$-dimensional conformal
algebra $o(d,2)$ one identifies the $AdS_{d}$ translations with
the mixture of translations and special conformal transformations
in the conformal algebra
\be
P^{a}_{AdS_{d}}=P^{a}-\lambda^{2}K^{a}\,.
\ee
The generators $P^{a}_{AdS_{d}}$ and $L_{ab}$ form the $AdS_{d}$
subalgebra $o(d-1,2)\subset o(d,2)$. This embedding breaks down
the manifest $o(1,1)$ dilatation covariance because it mixes the
operators $P^{a}$ and $K^{a}$, which have different dimensions.
$\lambda $ is the dimensionful Wigner-In\"{o}nu contraction
parameter to be identified with the inverse $AdS_d$ radius.

The $sp(2M)$ algebra admits analogous description  in terms of the
generators $L_{\al}{}^{\beta}, P_{\al\beta}, K^{\al\beta}$ and $
D$, where indices $\ga,\gb \ldots$  range from $1$ to $M$ and
$L_{\al}{}^{\beta}$ is traceless. The commutation relations are
\be\label{bCom}
 [K^{\alpha\beta},K^{\gamma\delta}]=0 \,,\qquad
[P_{\alpha\beta},P_{\gamma\delta}]=0\,,
\ee
\be
[D,P_{\alpha\beta}]=-P_{\alpha\beta} \,,\qquad
[D,K^{\alpha\beta}]=K^{\alpha\beta} \,,\qquad
[D,L_{\alpha}{}^{\beta}]=0\,,
\ee
\be
[L_{\alpha}{}^{\beta},P_{\gamma\delta}]=-\delta_{\gamma}
{}^{\beta}P_{\alpha\delta}
-\delta_{\delta}{}^{\beta}P_{\alpha\gamma}+\frac{2}{M}
\delta_{\alpha}{}^{\beta}P_{\gamma\delta}\,,
\ee
\be [L_{\alpha}{}^{\beta},
K^{\gamma\delta}]=\delta_{\alpha}{}^{\gamma}K^{\beta\delta}
+\delta_{\alpha}{}^{\gd}
K^{\beta\gamma}-\frac{2}{M}\delta_{\alpha}{}^{\beta}K^{\gamma\delta}\,,
\ee
\be
[P_{\al\gb},K^{\gamma\gd}]=L_{\al}{}^{\gd}\gd_{\gb}{}^{\gamma} +
L_{\gb}{}^{\gd}\gd_{\al}{}^{\gamma} +
L_{\al}{}^{\gamma}\gd_{\gb}{}^{\gd} +
L_{\gb}{}^{\gamma}\gd_{\al}{}^{\gd} + \f{4}{M}D( \gd_{\al}{}^{\gd}
\gd_{\gb}{}^{\gamma} +  \gd_{\gb}{}^{\gd} \gd_{\al}{}^{\gamma})\,,
\ee
\be\label{eCom}
[L_{\alpha}{}^{\beta},L_{\gamma}{}^{\delta}]=\delta_{\alpha}{}^{\delta}
L_{\gamma}{}^{
\beta}
-\delta_{\gamma}{}^{\beta}L_{\alpha}{}^{\delta}\,.
\ee
Note that the generalized Lorentz subalgebra generated by
$L_{\al}{}^{\beta}$ is $sl_M$. Analogously to the usual conformal
algebra, generalized translations generated by $P_{\ga\gb}$ form
Abelian subalgebra of $sp(2M)$. Generalized special conformal
transformations generate a dual Abelian subalgebra.

The commutation relations (\ref{bCom})-(\ref{eCom}) can be
realized by the vector fields
\be
P_{\alpha\beta}=\frac{\partial}{\partial X^{\alpha\beta}}\,,
\qquad K^{\alpha\beta}=4
X^{\alpha\gamma}X^{\beta\eta}\frac{\partial}{\partial
X^{\gamma\eta}}\,,
\ee
\be
L_{\alpha}{}^{\beta}=2 X^{\beta\gamma}\frac{\partial}{\partial
X^{\alpha\gamma}}-\f{2}{M} \gd_{\al}{}^{\gb}X^{\gb\gamma}\f{\p}{\p
X^{\gb\gamma}} \,,\qquad D=X^{\beta\gamma}\frac{\partial}{\partial
X^{\beta\gamma}}\,,
\ee
where $X^{\al\beta}=X^{\beta\al}$ are coordinates of $\M_M$.

The simplest way to see that the commutation relations
(\ref{bCom})-(\ref{eCom}) are indeed of $sp(2M)$ is to use its
oscillator realization \cite{BG}. Actually, let $\hat{a}_{\alpha}$
and $\hat{b}^{\alpha}$ be oscillators with the commutation
relations
\be
[\hat{a}_{\alpha},\hat{b}^{\beta}]=\delta_{\alpha}{}^{\beta}
\,,\qquad [\hat{a}_{\alpha},\hat{a}_{\beta}]=0 \,,\qquad
[\hat{b}^{\alpha},\hat{b}^{\beta}]=0\,.
\ee
The generators of $sp(2M)$ are spanned by the bilinears
\be
\hat{T}_{\alpha}{}^{\gb}
=\frac{1}{2}\{\hat{a}_{\alpha},\hat{b}^{\beta}\} \,,\qquad
\hat{P}_{\alpha\beta}=\hat{a}_{\alpha}\hat{a}_{\beta} \,,\qquad
\hat{K}^{\alpha\beta}=\hat{b}^{\alpha}\hat{b}^{\beta}\,.
\ee

Instead of working in terms of operators it is convenient to use
the star-product operation in the algebra of polynomials of
commuting variables $a_{\alpha}$ and $b^{\alpha}$
\be\label{old}
(f\star g)(a,b)=\frac{1}{\pi^{2M}}\int
f(a+u,b+t)g(a+s,b+v)e^{2(s_{\alpha}
t^{\alpha}-u_{\alpha}v^{\alpha})} \,
d^{M}u\,d^{M}t\,d^{M}s\,d^{M}v\,.
\ee
The star-product defined this way, often called Moyal product,
 describes the product of
symmetrized (i.e Weyl ordered) polynomials of oscillators in terms
of symbols of operators. The integral is normalized in such a way
that
\be
\frac{1}{\pi^{2M}}\int e^{2(s_{\alpha}
t^{\alpha}-u_{\alpha}v^{\alpha})} \,
d^{M}u\,d^{M}t\,d^{M}s\,d^{M}v=1 \,,
\ee
so that $1$ is the unit element of the algebra. Eq.(\ref{old})
defines the associative algebra with the defining relations
\be
[a_{\alpha},b^{\beta}]_{\star}= \delta_{\alpha}{}^{\beta}\,,\qquad
{[}a_{\alpha},a_{\beta}]_{\star}=0\,,\qquad
{[}b^{\alpha},b^{\beta}]_{\star}=0
\ee
($[a,b]_{\star} = a\star b - b\star a$). The star product
realization of the generators of $sp(2M)$ is
\be
T_{\alpha}{}^{\beta}=a_{\alpha}b^{\beta} \,,\qquad
P_{\alpha\beta}=a_{\alpha}a_{\beta} \,,\qquad
K^{\alpha\beta}=b^{\alpha}b^{\beta}\,,
\ee
where the $gl(M)$ generator $T_{\alpha}{}^{\beta}$ decomposes into
the $sl(M)$ ``Lorentz" and $o(1,1)$ ``dilatation" generators
\be
L_{\alpha}{}^{\beta}
=a_{\alpha}b^{\beta}-\frac{1}{M}\delta_{\alpha}{}^{\beta}a_{\gamma}b^{\gamma}
\,,\qquad D=\frac{1}{2}a_{\alpha}b^{\alpha}\,.
\ee
The bilinears of oscillators fulfil the commutation relations
(\ref{bCom})-(\ref{eCom}).

The embedding of the generalized $AdS$ subalgebra into the
conformal algebra $sp(2M)$ is achieved by identification of the
(generalized) $AdS$ translations with the mixture of translations
and special conformal transformations $
P^{AdS}_{\al\beta}=P_{\al\beta}+\lambda^{2}\eta_{\al\beta\gamma\delta}K^{\gamma\delta
} $ with some bilinear form $\eta_{\al\beta\gamma\delta}$. (Note
that keeping the same number of translation generators we  keep
dimension of the generalized space-time intact). As argued in
\cite{osp}, $\eta_{\al\beta\gamma\delta}$ has to have the
factorized form:
$\eta_{\al\beta\gamma\delta}=V_{\al\gamma}V_{\beta\delta}$, where
$V_{\al\beta}$ is some nondegenerate antisymmetric form (thus
requiring $M$ to be even). In what follows the form $V_{\ga\gb}$
will be used to raise and lower indices  according to the rule
\be
A_{\al}=V_{\beta\al}A^{\beta}\,,\qquad
A^{\al}=V^{\al\beta}A_{\beta}\,, \qquad
V_{\al\beta}V^{\al\gamma}=\delta_{\beta}{}^{\gamma}\,.
\ee
  Thus, the generalized $AdS$
translations have the form
\be\label{PAds}
P^{AdS}_{\al\beta}=P_{\al\beta}+\lambda^{2}V_{\al\gamma}
V_{\beta\delta}K^{\gamma\delta}=
P_{\al\beta}+\lambda^{2}K_{\al\beta}\,.
\ee
The commutation relations of $P^{AdS}_{\al\beta}$ have the form
 \be
 [P^{AdS}_{\al\beta},P^{AdS}_{\gamma\delta}]=2\lambda^{2}
 (V_{\beta\gamma}L^{AdS}_{\al
\delta}+V_{\beta\delta}
 L^{AdS}_{\al\gamma}+V_{\al\gamma}L^{AdS}_{\beta\delta}+
 V_{\al\delta}L^{AdS}_{\beta\gamma})\,,
 \ee
where $L^{AdS}_{\al\beta}= L^{AdS}_{\beta\al}$ are generators of
the $sp(M)$  subalgebra of $gl_M$ which leaves invariant the
symplectic form $V_{\ga\gb}$. The full generalized $AdS$
subalgebra is $sp(M)\oplus sp(M)\subset sp(2M)$. Its Lorentz
subalgebra $sp^{l}(M)$ identifies with the diagonal $sp(M)$ while
$AdS$ translations span $sp(M)\oplus sp(M)/sp^{l}(M)$. Note that
the generalized $dS$ algebra obtained from (\ref{PAds}) by virtue
of the sign change $\gl^{2}\to-\gl^{2}$ is $Sp(M,C)^{R}$.

\subsection{Fock space and $Sp(2M)$ covariant equations}

The $sp(2M)$ invariant equations of all massless fields in three
and four dimensions are naturally described \cite{Shayn,osp} in
terms of sections of the Fock fiber bundle over $\M_M$. In other
words, consider functions on $\M_M$ taking values in the Fock
module $F$
\be\label{Mod} |\Phi(b|X)\rangle=C(b|X)\star\Fock\,,
\ee
where $C(b|X)$ is some ``generating function''
\be\label{expans}
C(b|X)=\sum_{m=0}^{\infty}\frac{1}{m!}c_{\beta_{1}...\beta_{m}}
(X)b^{\beta_{1}}...b^{\beta_{m}}
\ee
and $\Fock$ is the Fock vacuum  defined by the relations
\be
a_{\al}\star\Fock=0 \,,\qquad \Fock\star b^{\al}=0\,.
\ee
$\Fock$ can be realized as an element of the star-product algebra
\be
\Fock=e^{-2a_{\al}b^{\al}}\,.
\ee
Note that the Fock vacuum is the space-time constant projector
\be\label{vac}
d\Fock=0\,,\qquad
 \Fock\star\Fock=\Fock\,,
\ee
where $d$ is de Rahm differential
\be
d=dX^{\ga\gb}\frac{\partial}{\partial X^{\ga\gb}} \,,\qquad
d^{2}=0.
\ee
As shown in \cite{osp} the relevant flat space
 $Sp(2M)$ covariant equation can be formulated in the form
\be\label{Fock}
d|\Phi(b|X)\rangle-w_{0}\star |\Phi(b|X)\rangle =0\,,
\ee
where
\be\label{go}
w_{0}=\frac{1}{2} dX^{\al\beta}a_{\al}a_{\beta} \,.
\ee
That the equation (\ref{Fock}) does indeed describe all conformal
field equations in $d=3$ and $d=4$ was shown in \cite{Shayn} and
\cite{osp} for the cases of $M=2$ and $M=4$, respectively. In this
paper we will consider the general case of any even $M$. It is
worth to mention that the cases
 of $M=8$, $M=16$, and
$M=32$ were argued  in \cite{Sp} to correspond to conformal
systems in $d=6$, $d=10$ and $d=11$, respectively.

The Fock fiber bundle realization of the higher spin equations
guarantees generalized conformal symmetry of the system along with
its infinite-dimensional higher spin extension.  Actually, let
$w_{0}$ be some 1-form, taking values in the higher spin algebra
identified with the star product algebra (i.e., the algebra of
regular functions of oscillators acting on the Fock module $F$)
\be
w_0 (X)=\sum_{m,n=0}^{\infty}\frac{1}{m!n!}
w_{_0\,\beta_{1}...\beta_{m}}{}^{\al_{1}...\al_{n}}
(X)a_{\al_{1}}...a_{\al_{n}}b^{\beta_{1}}...b^{\beta_{m}}\,,
\ee
which satisfies the zero-curvature condition
\be
\label{0cu} d w_{0}=w_{0}\star\wedge w_{0}\label{zero}\,.
\ee
The equations (\ref{Fock}), (\ref{zero}) are invariant under the
gauge transformations
\be
\delta w_0=d\epsilon-[w_0 ,\epsilon]_{\star}\,,
\ee
\be\label{trans}
\delta |\Phi(b|X)\rangle =\epsilon\star |\Phi(b|X)\rangle\,,
\ee
where $\epsilon(a,b|X)$ is an arbitrary infinitesimal gauge
parameter. Any fixed vacuum solution $w_{0}$ of the equation
(\ref{zero}) breaks the local higher spin symmetry to its
stability subalgebra with the infinitesimal parameters
$\epsilon_{0}(a,b|X)$ satisfying the equation
\be\label{eps}
d\epsilon_{0}-[w_{0},\epsilon_{0}]_{\star}=0.
\ee

Consistency of this equation is guaranteed by (\ref{zero}). As a
 result,
 (\ref{eps}) admits locally a pure gauge solution
\be\label{gdg}
w_{0}(X)=-g^{-1}(X)\star dg(X)\,,
\ee
where $g(a,b|X)$ is some invertible element of the star-product
algebra. The global symmetry parameters satisfying (\ref{eps})
then have the form
\be\label{epss}
\epsilon_{0}(X) = g^{-1}(X)\star\xi\star g(X)\,,
\ee
where an arbitrary $X$-independent element $\xi=\xi(a,b)$ of the
star-product algebra describes parameters of the global higher
spin symmetry which acts on the solutions of the equation
(\ref{Fock}) (for any given $w_0$). In particular, the $sp(2M)$
subalgebra spanned by bilinears of oscillators is thus shown to be
a symmetry of the equation (\ref{Fock}).

Analogously one solves the equation (\ref{Fock}) in the form
\be
\label{sol} |\Phi(b|X)\rangle=g^{-1}\star |\Phi(b|X_{0})\rangle
\,,
\ee
where $|\Phi(b|X_{0})\rangle$ plays a role of initial data. The
meaning of this formula is that the Fock module
$|\Phi(b|X_0)\rangle$ parametrises all combinations of the
derivatives of the dynamical fields at $X=X_0$ which are allowed
to be nonzero by the field equations. The formula (\ref{sol})
plays a role of  the covariantized Taylor expansion reconstructing
generic solution in terms of its derivatives
 at $X=X_0$. Note that the Fock module $F$ is not
unitary because it decomposes into an infinite sum of
finite-dimensional (tensor) representations of the generalized
noncompact Lorentz algebra $sl_M({\bf R})$. Nevertheless, the fact
that initial data of the problem are formulated in terms of the
Fock module $F$ is closely related to the fact (see e.g.
\cite{Fr,BLS}) that the collection of unitary massless
representations corresponding to this dynamical system in $d=4$ is
described by the unitary Fock module $U$ known as singleton
representation of $sp(8)$. (It is also well known that unitary
representations of the $4d$ conformal algebra associated with
massless fields admit Fock realization in terms of appropriate
oscillators \cite{Todor}.)
 As shown in \cite{Shayn,osp}, the modules $U$ and  $F$ are related
by some Bogolyubov transform.

The formulae (\ref{gdg}), (\ref{sol})  will play the key role in
our analysis. They allow one to solve the equations of motion
explicitly provided that the gauge function $g(X)$ is found that
corresponds to a chosen zero-curvature connection $w_0$. This
program for the flat connection (\ref{go}) was accomplished in
\cite{osp}. In this paper we will find a family of such gauge
functions $g(X)$ that all nonvanishing components of $w_0$ take
values in the $AdS$ subalgebra $osp(L|M)\oplus osp(L|M)$ of
$osp(2L|2M)$.

\section{$Sp(M)$ and star product}

As argued in \cite{osp}, the generalized $AdS$ space is identified
with $Sp(M)$. Let us note that the generalized conformal group
$Sp(2M)$ does not  act globally on $Sp(M)$ analogously to the
usual conformal group acting in the Minkowski space-time by
M\"{o}bius transformations which have singularities. Recall that
usual Minkowski space-time is the big cell of the compactified
Minkowski space. Analogously, the generalized Minkowski space-time
is the big cell in the compactified generalized space-time $\M_M$.
The universal covering space of $Sp(M)$ can be thought of as a
sort of deformation of the generalized Minkowski space-time being
the big cell of $\M_M$.

The group $Sp(M)$ is realized by the $M\times M$ matrices
$U_\ga{}^\gb$ satisfying
\be\label{Group}
U_{\al}{}^{\beta}U_{\gamma}{}^{\delta}V^{\al\gamma}=V^{\beta\delta}\,,
\ee
where $V^{\ga\gb}$ is some non-degenerate antisymmetric form
$V^{\ga\gb}=-V^{\gb\ga}$ ($M$ is even). The manifold $Sp(M)$ is
$\frac{M(M+1)}{2}$ dimensional. It can be described by local
coordinates $X^{\ga\gb}=X^{\gb\ga}$. The simplest parametrization
is
\be
\label{exp} U_{\al}{}^{\beta}=\left(\exp(\lambda
X)\right)_{\al}{}^{\beta}\,,
\ee
where $\lambda$ is inverse ``radius'' of $Sp(M)$ introduced to
compensate the space dimensionality of  $X^{\al\beta}$. Note that
a particular value of $\lambda \neq 0$ is irrelevant unless there
are some other dimensionful parameters in the theory (e.g., the
gravitational constant). The exponential in (\ref{exp}) is the
matrix exponential of
\be
X_\ga{}^\gb = V_{\gga\ga} X^{\gga\gb}\,.
\ee
It is elementary to see that the parametrization (\ref{exp})
solves the group equation (\ref{Group}). The exponential
parametrization (\ref{exp}) provides the universal covering space
\cite{Barut} of $Sp(M)$ (metaplectic group $Mp(M)$) topologically
equivalent to $R^{\f{M(M+1)}{2}}$, the big cell of $\M_M$.

$Sp(M)$ is invariant under the action of $Sp(M)\times Sp(M)$
generated by left and right actions of $Sp(M)$ on itself. Using
the oscillator realization of $sp(M)\oplus sp(M) \subset sp(2M)$
we can set
\be
\label{stacon} w_{0}(X) =\omega_{\al\beta}(X) a^{\al}b^{\beta}
+h_{\al\beta}(X) (a^{\al}a^{\beta} +\gl^{2}b^{\al}b^{\beta})\,,
\ee
where the ``Lorentz connection'' $\omega_{\al\beta}(X)$ and the
``frame'' $h_{\al\beta}(X)$ have the form
\be\label{Lor}
\omega_{\al\beta}=-\frac{1}{2}\left(d(U^{-1})_{\al}{}^{\gamma}U_{\gamma\beta}+
dU_{\al}{}^{\gamma}(U^{-1})_{\gamma\beta}\right)\,,
\ee
\be\label{frame}
h_{\al\beta}=\frac{1}{4\gl}\left(dU_{\al}{}^{\gamma}(U^{-1})_{\gamma\beta}-
d(U^{-1})_{\al}{}^{\gamma}U_{\gamma\beta}\right)\,,
\ee
which guarantees that $w_{0}$ satisfies (\ref{0cu}). In the
exponential parametrization (\ref{exp}) one gets
\be
\label{lorcon}
\omega^{\al\beta}=\frac{\lambda}{2}dX_{\mu\nu}\left(
\int\limits_{0}^{1}\exp(\lambda Xt)^{\mu\gb}\exp(\lambda
Xt)^{\nu\al}dt-\int\limits_{-1}^{0}\exp(\lambda
Xt)^{\mu\gb}\exp(\lambda Xt)^{\nu\al}dt\right)\,,
\ee
\be\label{frameX}
h^{\al\beta}=
\frac{1}{4}dX_{\mu\nu}\int\limits_{-1}^{1}\exp(\lambda
Xt)^{\mu\gb}\exp(\lambda Xt)^{\nu\al}dt\,,
\ee
where we used  the identity $\delta
e^{A}=\int\limits_{0}^{1}e^{At}\delta A e^{A(1-t)}dt$ valid
 for an arbitrary matrix $A$. As expected, in
 the flat limit $\lambda\to 0$ one recovers  (\ref{go}).

Let us now present the star product pure gauge form  (\ref{gdg})
of the connection (\ref{stacon})-(\ref{frame}). The final result
is
\be\label{main}
g=\f{1}{\det\left\|ch\frac{\lambda X}{2}\right\|} \exp\Big
(-\frac{1}{\lambda}\left(th\frac{\lambda
X}{2}\right)^{\al\beta}(a_{\al}a_{\beta}+\lambda^{2}b_{\al}b_{\beta})\Big)\,,
\ee
\be
g^{-1}=\f{1}{\det\left\|ch\frac{\lambda X}{2}\right\|} \exp\Big
(\frac{1}{\lambda}\left(th\frac{\lambda
X}{2}\right)^{\al\beta}(a_{\al}a_{\beta}+\lambda^{2}b_{\al}b_{\beta})\Big)\,.
\ee

This formula is derived as follows. Let $sp(M)$ be realized in
terms of bilinears of oscillators $\ga_\ga$ satisfying the
commutation relations
\be
[\al_{\al},\al_{\beta}]_{*}= -2V_{\al\beta}\,,
\ee
with the star-product
\be\label{new} (f*g)(\al)=\frac{1}{(2\pi)^{M}}\int f(\alpha+u)
g(\alpha +v)e^{u_{\alpha}v^{\alpha}} \,d^{M}u \,d^{M}v \,.
\ee

Consider star-product algebra elements $g_{1}$ and $g_{2}$ of the
form
\be
g_{1}=r_{1}e^{\frac{1}{2} f_{1}^{\al\beta}\al_{\al}\al_{\beta}}
\,,\qquad g_{2}=r_{2}e^{\frac{1}{2}
f_{2}^{\al\beta}\al_{\al}\al_{\beta}}\,.
\ee
with some $\al$-independent $r_{1}$, $r_{2}$, $f^{\ga\gb}_{1}$ and
$f^{\ga\gb}_{2}$. Elementary evaluation of the Gaussian integrals
shows that
\be g_{1,2}=g_{1} * g_{2}=
r_{1,2}e^{\frac{1}{2} (f_{1}\circ
f_{2})^{\al\beta}\al_{\al}\al_{\beta}}\,,
\ee
where
\be
r_{1,2} =
\frac{r_{1}r_{2}}{\sqrt{\det\left\|f_{1}f_{2}+1\right\|}}
\ee
and
\be
\label{fof} f_{1}\circ
f_{2}=\frac{1}{1+f_{2}f_{1}}(1+f_{2})-\frac{1}{1+f_{1}f_{2}}(1-f_{1})
\ee
(with the  usual matrix multiplication on the right hand side:
$AB\to A_{\al}{}^{\gamma}B_{\gamma}{}^{\beta}$, $\frac{1}{A}B\to
(A^{-1})_{\al}{}^{\gamma}B_{\gamma}{}^{\beta}$). Let us look for a
map
\be
\label{hom} g(U)=r(U)e^{\frac{1}{2}
f^{\al\beta}(U)\al_{\al}\al_{\beta}}
\ee
of $Sp(M)$ into the star-product algebra, such that
\be\label{comp}
g(U_{1}) *  g(U_{2})= g(U_{1}U_{2})=r(U_{1}U_{2})e^{\frac{1}{2}
f^{\al\beta}(U_{1}U_{2})\al_{\al}\al_{\beta}}\,.
\ee
Equivalently, one can use the inverse map $U(f)$ requiring
\be\label{eq}
U(f_{1})U(f_{2})=U(f_{1}\circ f_{2}) \,.
\ee

As shown in appendix, the multiplication law (\ref{fof}) requires
 \be\label{Gom}
U^{\al\beta}(f) =\left(\frac{1+f}{1-f}\right)^{\al\beta} \,.
\ee
 The inverse formula is analogous
\be\label{Gom1}
f^{\al\beta}(U) =\left(\frac{U-1}{1+U}\right)^{\al\beta}\,.
\ee
The normalization factor is
\be\label{factor}
r(U)=\frac{2^{\frac{M}{2}}}{\sqrt{\det\left\|U+1\right\|}}\,.
\ee

To derive (\ref{main}) it remains to use (\ref{exp}) and to
observe that the two $sp(M)$ subalgebras of $sp(2M)$ are generated
by the two mutually commuting sets of oscillators
\be
\al_{\al}^{\pm}=\frac{a_{\al}}{\sqrt{\gl}}\pm\sqrt{\gl}
V_{\beta\al}b^{\beta}=\frac{1}{\sqrt{\gl}}(a_{\al}\pm\gl
b_{\al})\,,
\ee
satisfying the commutation relations
\be
[\al^{\pm}_{\al},\al^{\pm}_{\beta}]_{*}=\pm 2V_{\al\beta}\,.
\ee

The map (\ref{Gom}) has a number of interesting properties. In
particular,
\be\label{anamor}
U^{-1}(f)=U(-f)\,,
\ee
\be\label{inver}
U(f)=-U^{-1}(-f^{-1})\,.
\ee
 The property (\ref{anamor}) is a
consequense of the elementary fact (see e.g. \cite{Vas}) that the
star product  (\ref{new}) admits an antiautomorphism $\rho(g(\al))
= g(i \al)$, i.e. $\rho (g_1) * \rho (g_2 )$ = $\rho (g_2 * g_1
)$.  {}From (\ref{hom}) it follows that $\rho (U (f) ) = U(-f)$.
The natural group antiautomorphism is $\rho (U) = U^{-1}$. The
formula (\ref{anamor}) identifies the antiautomorphism $\rho$ in
the star product algebra with that of the group $Sp(M)$.

The formula (\ref{inver}) is more interesting. It does not have a
global interpretation within $Sp(M)$ being singular at degenerate
$f^{\ga\gb}$ (in particular for $f^{\ga\gb}=0$ and, therefore,
$U=I$). However, these maps are expected to have global meaning in
$\M_M$ where one can define inversion by analogy with the flat
case considered in \cite{Sp}
\be
\label{inv} I(f) = -f^{-1}\,,\qquad I(U) = - U^{-1}\,.
\ee
The formula (\ref{inver}) implies that these two definitions are
consistent with each other. Note that inversion defined this way
maps unit element of $Sp(M)$ to the central element $-I$ which
does not belong to the connected component of unity $PSp(M)
\subset Sp(M)$.

\section{Arbitrary coordinates}
\label{Arbitrary coordinates}

The gauge function (\ref{main}) corresponds to the exponential
realization of $Sp(M)$, thus yielding global coordinates which
cover the metaplectic group $Mp(M)$. Our formalism allows one to
write down explicit form of vacuum gauge connections (Cartan
forms) in arbitrary coordinates, however. Indeed, let us consider
a gauge function of the form
\bee
g=\sqrt{\det\left\|1-\gl^{2} f^{2}(X)\right\|
}\exp\Big(-f(X)^{\al\gb}(a_{\al}a_{\gb}+
\gl^{2}b_{\al}b_{\gb})\Big)\,,\nn\\
g^{-1}=\sqrt{\det\left\|1-\gl^{2} f^{2}(X)\right\|
}\exp\Big(f(X)^{\al\gb}(a_{\al}a_{\gb}+
\gl^{2}b_{\al}b_{\gb})\Big)\,,
\eee
where $f^{\al\gb}(X)=f^{\gb\al}(X)$ is an arbitrary function of
matrix coordinates $X^{\al\gb}$. The zero-curvature connection
(\ref{gdg}) can be written in the form
\be
w_0=-g(-f)\star \Big (dX^{\al\gb}\f{\p f_{1}^{\gamma\gl}}{\p
X^{\al\gb}}\f{\p}{\p f_{1}^{\gamma\gl}} g(f_{1})\Big )\Big
|_{f_{1}=f}\,.
\ee
Direct computation leads to the expressions for the ``Lorentz
connection" and  ``frame"
\be
\label{geh} h^{\al\gb}=dX^{\rho\gs}\Big (\f{1}{1-\gl^{2}f^{2}}\Big
)^{\al\gamma} \Big (\f{\p{f_{\gamma}{}^{\gl}}}
{\p{X^{\rho\gs}}}-\gl^{2} f_{\gamma}{}^{\mu}\f{\p f_{\mu}{}^{\nu}}
{\p X^{\rho\gs}} f_{\nu}{}^{\gl}\Big )\Big
(\f{1}{1-\gl^{2}f^{2}}\Big )_{\gl}{}^{\gb}\,,
\ee
\be
\label{gego} \omega^{\al\gb}=2\gl^{2}dX^{\rho\gs}\Big
(\f{1}{1-\gl^{2}f^{2}}\Big )^{\al\gamma}\Big (\f{\p
f_{\gamma}{}^{\mu}}{\p X^{\rho\gs}} f_{\mu}{}^{\gl}-
f_{\gamma}{}^{\mu}\f{\p f_{\mu}{}^{\gl}}{\p X^{\rho\gs}}\Big )
\Big (\f{1}{1-\gl^{2}f^{2}}\Big )_{\gl}{}^{\gb}\,.
\ee
Note that from these formulae it follows that
\be\label{w(h)}
\omega^{\al\gb}=2h^{\al\gamma}f_{\gamma}{}^{\gb}-2f^{\al\gamma}
h_{\gamma}{}^{\gb}+\gl^{2}f^{\al\gamma}\omega_{\gamma}{}^{\gl}
f_{\gl}{}^{\gb}\,.
\ee

An arbitrary function $f^{\ga\gb} (X)$ parametrizes various
coordinate choices in $Sp(M)$. A relationship with the coordinates
of the exponential parametrization obviously is
\be
f(\tilde{X}) = \frac{1}{\lambda} \,th \frac{\lambda X}{2}
\ee
which implies locally
\be
\label{map} sh \frac{\lambda X}{2} = \frac{\lambda
f(\tilde{X})}{\sqrt{1-\lambda^2 f^2(\tilde{X})}}\,,\qquad ch
\frac{\lambda X}{2} = \frac{1}{\sqrt{1-\lambda^2
f^2(\tilde{X})}}\,.
\ee

The formulae (\ref{geh}) and (\ref{gego}) thus provide a
representation for Cartan forms in arbitrary coordinates
associated with one or another function  $f^{\ga\gb} (X)$.
Consider now a few particular examples. Let
 $f^{\al\gb}(X)$ be of the form
\be
f^{\al\gb}(X)=\phi(\det\|X\|)X^{\al\gb}\,.
\ee
The corresponding connections are
\bee
h^{\al\gb}=\phi\cdot \Big (\f{1}{1-\gl^{2}\phi^{2}X^{2}} \Big
)^{\al\gamma}(dX_{\gamma}{}^{\gl}-\gl^{2}\phi^{2}X_{\gamma}{}^{\mu}
dX_{\mu}{}^{\nu}X_{\nu}{}^{\gl})\Big
(\f{1}{1-\gl^{2}\phi^{2}X^{2}} \Big )_{\gl}{}^{\gb}+\nn\\
+\tilde{\phi}\cdot dX^{\rho\gs}(X^{-1})_{\rho\gs}\Big
(\f{X}{1-\gl^{2}\phi^{2}X^{2}}\Big )^{\al\gb}\,,
\eee
\be
\omega^{\al\gb}=2\gl^{2}\phi^{2}\cdot \Big
(\f{1}{1-\gl^{2}\phi^{2}X^{2}} \Big
)^{\al\gamma}(dX_{\gamma}{}^{\mu}X_{\mu}{}^{\gl} -
X_{\gamma}{}^{\mu}dX_{\mu}{}^{\gl})\Big
(\f{1}{1-\gl^{2}\phi^{2}X^{2}} \Big )_{\gl}{}^{\gb}\,,
\ee
where
\be
\tilde{\phi}=\f{\p\phi}{\p\ln{\det\|X\|}}\,.
\ee

Another useful example results from
\be
f^{\pm}_{\al\gb} (X)=\Big (\f{X}{1 \mp \sqrt{1 - \gl^{2} X^{2}}}
\Big )_{\al\gb}\,.
\ee
The corresponding gauge function is
\be\label{stereo}
g^{\pm}=\det\left\|\frac{1+\gl X\pm\sqrt{1-\gl^{2}X^{2}}}{\gl
X}\right\|\exp\left(-\left(\frac{X}{1 \mp
\sqrt{1-\gl^{2}X^{2}}}\right)^{\al\gb}(a_{\al}a_{\gb}+\gl^{2}
b_{\al}b_{\gb})\right)\,.
\ee
In these ``stereographic'' coordinates the ``frame" gets the
following simple form
\be\label{Stframe}
h^{\al\gb}=\f{1}{2}\Big (\f{1}{\sqrt{1-\gl^{2}X^{2}}}\Big
)^{\al\gamma}dX_{\gamma}{}^{\gl}\Big
(\f{1}{\sqrt{1-\gl^{2}X^{2}}}\Big )_{\gl}{}^{\gb}\,.
\ee
Let us now compare this formula with those obtained in
\cite{Shayn,Bol} to describe massless fields in $AdS_{3}$($M=2$)
and $AdS_{4}$($M=4$).

Let us first consider the case $M=2$. Using for example $g^{+}$,
from (\ref{stereo}) one obtains
\be\label{AdS3}
g=\frac{2\sqrt{z}}{1+\sqrt{z}}\exp\left(-\frac{1}{1+\sqrt{z}}x^{\al\beta}
(a_{\al}a_{\beta}+\lambda^{2}b_{\al}b_{\beta})\right)\,,
\ee
\be
g^{-1}=\frac{2\sqrt{z}}{1+\sqrt{z}}\exp\left(\frac{1}{1+\sqrt{z}}x^{\al\beta}
(a_{\al}a_{\beta}+\lambda^{2}b_{\al}b_{\beta})\right)\,,
\ee
where $z=1+\frac{1}{2}\lambda^{2}x_{\al\beta}x^{\al\beta}$. The
``frame" and the ``Lorentz connection" are
\be
\label{hgo3} h_{\al\beta}=\frac{1}{2z}dx_{\al\beta}\,,\qquad
\omega_{\al\beta}=\frac{1}{2z}(dx_{\al}{}^{\gamma}x_{\gamma\beta}+
dx_{\beta}{}^{\gamma}x_{\gamma\al}).
\ee
To derive this result, which reproduces that of \cite{Shayn}, we
used  a simple fact that, when $M=2$, any antisymmetric matrix is
proportional to $V_{\al\beta}$ and, therefore, any polynomial of
matrix coordinates $P(x)^{\al\beta}$ decomposes into a combination
of its symmetric part $P_{S}(x_{\mu\nu}x^{\mu\nu})x^{\al\beta}$
and antisymmetric part $P_{A}(x_{\mu\nu}x^{\mu\nu})V^{\al\beta}$.
{}From (\ref{hgo3}) it follows that the metric tensor is
\be
g_{mn}=\frac{1}{2}h_{\al\beta ,n}h^{\al\beta}{}_{,m}=
\frac{1}{4}\frac{\eta_{mn}}{(1+\lambda^{2}x_{k}x^{k})^{2}}\,,
\ee
where
\be
x_{n}=\sigma_{n}{}^{\al\beta}x_{\al\beta} \,,\qquad
x_{\al\beta}=\frac{1}{2}\sigma_{\al\beta}{}^{n}x_{n}\,,
\ee
and  $\sigma_{n}{}^{\al\beta}$ is a set of basis symmetric real
matrices normalized to satisfy
\be
\sigma_{n}{}^{\al\beta}\sigma_{m}{}_{\al\beta}=2\eta_{mn}\,,
\ee
where $\eta_{mn}$ is the flat Minkowski metric.

To consider the $4d$ case we embed $AdS_{4}$ space-time into
$\M_4$ as follows
\be\label{emb}
X^{\al\beta}=\left(
\begin{array}{cc}
\mathbf{0} & x^{\ab\dbb}\\
x^{\bb\dab} & \mathbf{0}
\end{array} \right)\,,
\ee
where $\ab,\bb=1,2$, $\dab,\dbb=3,4$, and $x^{\ab\dbb}$ are local
$AdS_{4}$ coordinates which can be expressed via the vector
coordinates $x^{n}$ ($n=0\dots 3$) with the aid of Pauli matrices
$\gs_{n}{}^{\ab\dbb}=(I,\gs_{1}{}^{\ab\dbb}\dots\gs_{3}{}^{\ab\dbb})$
as
\be
x^{n}=\gs^{n}{}_{\ab\dbb}x^{\ab\dbb} \,,\quad
x^{\ab\dbb}=\frac{1}{2}x_{n}\gs^{n}{}^{\ab\dbb} \,,\quad
\gs_{n}{}_{\ab\dbb}\gs_{m}{}^{\ab\dbb}=2\eta_{mn}\,.
\ee
The  gauge function and gravitational fields resulting from
 (\ref{stereo}) and (\ref{Stframe}) are
\be\label{AdS4}
g=\Big(\frac{2\sqrt{z}}{1+\sqrt{z}}\Big)^{2}
\exp\left(-\frac{1}{1+\sqrt{z}}x^{\ab\dbb}
(a_{\ab}a_{\dbb}+\lambda^{2}b_{\ab}b_{\dbb})\right)\,,
\ee
\be
h_{\ab\dbb}=\frac{1}{2z}dx_{\ab\dbb}\,,
\ee
\be
\omega_{\ab\bb}=\frac{1}{2z}(dx_{\ab}{}^{\dbg}x_{\bb\dbg}+dx_{\bb}
{}^{\dbg}x_{\ab\dbg })\,,\qquad
\bar{\omega}_{\dab\dbb}=\frac{1}{2z}(dx^{\dg}{}_{\dab}x_{\dg\dbb}+
dx^{\dg}{}_{\dbb}x_{\dg\dab})\,,
\ee
where
$z=1+\frac{1}{2}\lambda^{2}x_{\ab\dbb}x^{\ab\dbb}=1+\gl^{2}x_{n}x^{n}$.
These $AdS_4$ gravitational fields coincide with those found in
\cite{Bol}.

\section{Symmetries}
Having fixed some vacuum solution $w_{0}$ of (\ref{zero}), the
local higher spin symmetry is broken down to the global one with
the parameter $\epsilon_{0}(a,b|X)$ satisfying (\ref{eps}). Once
the vacuum solution $w_{0}$ is fixed in the pure gauge form
(\ref{gdg}) with some gauge function $g$, it is easy to find the
gauge parameter $\epsilon_{0}(a,b|X)$ of the leftover global
symmetry. Indeed let the generating parameter $\xi (a,b;\mu,\eta)$
in (\ref{epss}) be of the form
\be
\xi=\xi_{0}\exp(a_{\al}\mu^{\al}-b^{\al}\eta_{\al})\,,
\ee
where $\xi_{0}$ is an infinitesimal constant while $\mu^{\al}$ and
$\eta_{\al}$ are constant parameters. An arbitrary symmetry with
star-product polynomial parameters can be obtained via
differentiation of $\xi$ with respect to $\mu^{\al}$ and
$\eta_{\al}$. Substitution of (\ref{main}) into (\ref{epss}) gives
\be\label{gauge}
\epsilon_{0}(a,b;\mu,\eta|X)=g^{-1}\star\xi\star g=
\xi_{0}\exp(a_{\al}\hat{\mu}^{\al}-b^{\al}\hat{\eta}_{\al})\,,
\ee
where
\be
\hat{\mu}^{\al}=ch(\gl X)^{\al\gb}\mu_{\gb}-\frac{sh(\gl
X)^{\al\gb}}{\gl}\eta_{\gb}\,,\quad \hat{\eta}^{\al}=ch(\gl
X)^{\al\gb}\eta_{\gb}-\gl\cdot sh(\gl X)^{\al\gb}\mu_{\gb}.
\ee
According to (\ref{map}), in the arbitrary coordinates associated
with the function $f^{\ga\gb} (X)$ of section \ref{Arbitrary
coordinates}, we have
\be
\hat{\mu}_{\al}=\Big (\f{1+\gl^{2}f^{2}(X)}{1-\gl^{2}f^{2}(X)}\Big
)_{\al}{}^{\gb}\mu_{\gb}-\Big (\f{2f(X)}{1-\gl^{2}f^{2}(X)}\Big
)_{\al}{}^{\gb}\eta_{\gb}\,,
\ee
\be
\hat{\eta}_{\al}=\Big
(\f{1+\gl^{2}f^{2}(X)}{1-\gl^{2}f^{2}(X)}\Big
)_{\al}{}^{\gb}\eta_{\gb}-\gl^{2}\Big
(\f{2f(X)}{1-\gl^{2}f^{2}(X)}\Big )_{\al}{}^{\gb}\mu_{\gb}\,.
\ee

The global symmetry transformation of the higher spin generating
function
\be
\delta |\Phi (b|X)\rangle \equiv \epsilon_0 \star |\Phi
(b|X)\rangle =\xi_{0}
\exp(-\hat{\eta}_{\al}b^{\al}+\frac{1}{2}\hat{\eta}^{\al}\hat{\mu}_{\al})
\cdot C(b+\hat{\mu}|X) \star\Fock\,
\ee
implies
\be\label{transC}
\delta C(b|X) = \xi_{0}C(b+\hat{\mu}|X)\exp \Big (\f{1}{2}
\hat{\eta}^{\al}\hat{\mu}_{\al}-b^{\al}\hat{\eta}_{\al}\Big )\,.
\ee
The  dynamical fields are associated with the scalar $c(X)=
C(0|X)$ and svector $c_\ga (X)=\frac{\partial}{\partial b^\al}
C(b|X)\Big |_{b=0}$ in the expansion (\ref{expans}). (All other
fields in $C(b|X)$ are expressed via derivatives of the dynamical
fields \cite{osp}.) Their transformation laws are
\be
\gd c(X)=\xi_{0}C(\hat{\mu}|X)\exp(\f{1}{2}
\hat{\eta}^{\al}\hat{\mu}_{\al})\,,
\ee
\be
\gd c_{\al}(X)=\xi_{0}\Big ( \f{\p}{\p{b^{\al}}} C(b+\hat{\mu}|X)
\Big|_{b=0} - \hat{\eta}_{\al} C(\hat{\mu}|X) \Big ) \exp(\f{1}{2}
\hat{\eta}^{\al}\hat{\mu}_{\al})\,.
\ee
 Differentiating over the parameters
$\mu^\ga$ and $\eta_\ga$ and setting them then equal to zero one
obtains explicit expressions for the higher spin symmetry
transformations associated with any symmetry parameters
$\epsilon_{0}(a,b|X)$ polynomial in the oscillators $a$ and $b$.
In particular, the transformation law with the parameters bilinear
in the oscillators reproduces the $Sp(2M)$ generalized conformal
transformations in the generalized $AdS$ space-time $Sp(M)$.

\section{Light-like solutions}
Once the gauge function $g$ is known one solves the system of free
field equation (\ref{Fock}) for all massless fields via
(\ref{sol}). Let us consider basis light-like solutions generated
by the initial data of the form
\be
C(b|0)=C_{0}\exp(\kappa_{\al}b^{\al})\,,
\ee
where $C_{0}$ is an arbitrary constant and $\kappa^{\al}$ is some
space-time constant spinor. According to (\ref{Mod}) the Fock
representation of the initial data has the form
\be
|\Phi(b|0)\rangle=C_{0}\exp(\kappa_{\al}b^{\al})\star
|0\rangle\langle 0|.
\ee
So the dynamical problem is solved by
\be
\label{solphi} |\Phi(b|X)\rangle=g^{-1}(X)\star
|\Phi(b|0)\rangle=\f{C_{0}}{\det\left\|ch\frac{\lambda
X}{2}\right\|} e^{\frac{1}{\lambda}\left(th\frac{\lambda
X}{2}\right)^{\al\beta}(a_{\al}a_{\beta}+\lambda^{2}b_{\al}b_{\beta})}
\star e^{\kappa_{\al}b^{\al}}\star e^{-2a_{\al}b^{\al}}\,.
\ee
Elementary evaluation of Gaussian integrals gives the following
result
\be
C(b|X)=\frac{C_{0}}{\sqrt{\det\left\|ch\lambda X\right\|}}
\exp\Big(t^{\al\gb}(\gl^{2}b_{\al}b_{\gb}+\kappa_{\al}\kappa_{\gb})+
p_{\gb}{}^{\al} \kappa_{\al}b^{\gb}\Big)\,,
\ee
where we use notations
\be
\begin{array}{c}
t_{\al}{}^{\beta}=\Big( \f{th(\gl X)}{2\gl}\Big )_{\al}{}^{\gb}
\,,\qquad p_{\al}{}^{\beta}=(ch^{-1}(\gl X))_{\al}{}^{\beta}\,,
\end{array}
\ee
equivalent by virtue of (\ref{map}) to
\be
t_{\al}{}^{\gb}=\Big (\f{f(X)}{1+\gl^{2}f^{2}(X)}\Big
)_{\al}{}^{\gb}\,,\qquad p_{\al}{}^{\gb}=\Big
(\f{1-\gl^{2}f^{2}(X)}{1+\gl^{2}f^{2}(X)}\Big )_{\al}{}^{\gb}\,.
\ee

Let us stress that, according to \cite{osp,Sp}, for the particular
case of $M=4$ the obtained expressions describe solutions of
massless equations for all spins in $AdS_4$, constructed in
\cite{Bol}. Using (\ref{AdS4}), these solutions take the form
\be
C(b|x)=z\exp\Big(\f{x^{\ab\dbb}}{2}(\kappa_{\ab}\kappa_{\dbb}+
\lambda^{2}b_{\ab}b_{\dbb})+\sqrt{z}\kappa_{\al}b^{\al}\Big)\,.
\ee
For the case of $M=2$ we get solutions of the $AdS_3$ massless
equations discussed in \cite{Shayn} of the form
\be
C(b|x)=\sqrt{z}\exp\Big(\f{x^{\al\beta}}{2}(\kappa_{\al}\kappa_{\beta}+
\lambda^{2}b_{\al}b_{\beta})+\sqrt{z}\kappa_{\al}b^{\al}\Big)\,.
\ee
Here we make use of the gauge function (\ref{AdS3}).

For the dynamical fields we obtain
\be
c(X)=C_{0}\sqrt{\det\left\|\frac{1-\gl^{2}f^{2}(X)}
{1+\gl^{2}f^{2}(X)}\right\|}
\exp(t^{\al\beta}\kappa_{\al}\kappa_{\beta}) \,,\qquad
\ee
\be
c_{\al}(X)=C_{0}\sqrt{\det\left\|\frac{1-\gl^{2}f^{2}(X)}
{1+\gl^{2}f^{2}(X)}\right\|}
p_{\al}{}^{\gb}\kappa_{\beta}\exp{(t^{\al\beta}\kappa_{\al}\kappa_{\beta})}\,.
\ee

Substitution of $\epsilon_{0}$ into (\ref{trans})  gives the
global higher spin symmetry transformation of the solution
(\ref{solphi})
\bee
\gd C(b|X)= C_{0}\xi_{0}\sqrt{\det\left\|\frac{1-\gl^{2}f^{2}(X)}
{1+\gl^{2}f^{2}(X)}\right\|}\exp\Big( t^{\al\beta}\gl^{2}(
b_{\al}+\hat{\mu}_{\al}) (b_{\gb}+\hat{\mu}_{\gb})+
\nn\\
 t^{\al\gb}\kappa_{\al}\kappa_{\gb}+p_{\gb}{}^{\al}\kappa_{\al}(b^{\gb} +
\hat{\mu}^{\gb})-\hat{\eta}_{\al}(b^{\al}+\frac{1}{2}\hat{\mu}^{\al})\Big)\,.
\eee

The flat limit $\lambda \to 0$ gives
\bee
\gd C(b|X)=C_{0}\xi_{0}\exp\Big(\f{1}{2}X^{\al\beta}(\kappa_{\al}
\kappa_{\gb}-2\kappa_{\al}\eta_{\gb}+\eta_{\al}\eta_{\gb})+\nn\\
+b^{\al}(\kappa_{\al}-\eta_{\al})+\kappa_{\al}\mu^{\al}+\frac{1}{2}
\mu_{\al}\eta^{\al}\Big).
\eee
For the dynamical fields we get the plane wave solutions
\be
c^{plane}(X)=C_{0}e^{\f{1}{2}X_{\al\beta}\kappa^{\al}\kappa^{\beta}}
\,,\qquad c^{plane}_{\al}(X)=C_{0}\kappa_{\alpha}
e^{\f{1}{2}X_{\al\beta}\kappa^{\al}\kappa^{\beta}}\,,
\ee
with the twistorial ``wave vector''
$K_{\al\beta}=\f{1}{2}\kappa_{\al}\kappa_{\beta}$.

The solution deformed to the $AdS$ case is not strictly speaking
plane wave. However it is ``conformally plane wave'' in the sense
that it has still leftover generalized conformal invariances
identified with such global symmetry transformations that
\be
\delta_{\epsilon_{0}}|\Phi(b|X)\rangle=0\,.
\ee
It is easy to see using (\ref{gauge}) that this condition is
solved by any parameter of the form
\be
\xi=f(a,b)\star(\rho^{\al}a_{\al})\,,
\ee
where $\rho$ is an arbitrary parameter such that
$\rho^{\al}\kappa_{\al}=0$ and $f(a,b)$  is an arbitrary function.
Indeed, according to (\ref{sol})
\be
\delta_{\epsilon_{0}}|\Phi(b|X)\rangle=g^{-1}\star\xi\star
C(b|0)\star\Fock= g^{-1} \star f \star (\rho^{\al}a_{\al})\star
e^{\kappa_{\al}b^{\al}}\star\Fock=0\,.
\ee

\section{Superextension}
The star-product formalism we use admits a straightforward
generalization to the supersymmetric case associated with $OSp(\L
|2M)$ where $\L$ is an arbitrary integer. To describe $osp(L|2M)$
superalgebra let us introduce the Clifford elements $\psi_{i}$
$(i=1\dots L)$ satisfying the anticommutation relations
\be\label{acom}
\{\psi_{i},\psi_{j} \}_{*}=\eta_{ij}\,,
\ee
where $\eta_{ij}=\eta_{ji}$ is some non-degenerate symmetric form.
The Clifford star-product in (\ref{acom}) is defined (see, e.g.,
\cite{Vas}) according to
\be
(f*g)(\psi)=\f{1}{2^{L}}\int
f(\psi+\phi)g(\psi+\chi)e^{-2\chi^{i}\phi_{i}} d^{L}\phi d^{L}\chi
\,,
\ee
where $\chi_{i}$ and $\phi_{i}$ are anticommuting variables. The
supercharges
\be
Q_{i\al}=a_{\al}\psi_{i}\,, \qquad S^{\al}_{i}=b^{\al}\psi_{i}
\ee
satisfy
\be
\{ Q_{i\al}, Q_{i\gb}\}_*=\eta_{ij}P_{\al\gb}\,, \qquad \{
S^{\al}_{i}, S^{\gb}_{j}\}_*=\eta_{ij}K^{\al\gb}\,.
\ee
Let the Grassmann odd coordinates $\theta^{i\al}$ be associated
with the $Q$-supergenerators. It is convenient to require the
differential $d\theta^{i\al}$ to anticommute to $dX^{\al\beta}$
and $\theta^{i\al}$.

It is easy to see \cite{osp} that the gauge function
\be
g=e^{-X^{\al\gb}a_{\al}a_{\gb}-\theta^{i\al}a_{\al}\psi_{i}}
\ee
reproduces the flat superspace  vacuum 1-form
\be
w_{0}=\Big(dX^{\al\gb}+\f{1}{2}d\theta^{i\al} \theta_{i}{}^{\gb}
\Big)P_{\al\gb}+d\theta^{i\al}Q_{i\al}\,.
\ee

The left Fock module $|\Phi(b,\psi^{+}|X,\theta)\rangle$ satisfies
the $osp(L|2M)$ supersymmetric equation
\be (d-w_{0})\star
|\Phi(b,\psi^{+}|X,\theta)\rangle=0\,,
\ee
where the supersymmetric Fock vacuum $\Fock$ in addition to
(\ref{vac}) is annihilated by the $\f{1}{2}L$ (in case of even
$L$) or $\f{1}{2}(L-1)$ (in case of odd $L$) annihilation Clifford
elements $\psi^{-}$ and, when L is odd, it is an eigenvector of
the central element $\Psi^{L}=\psi_{1}\dots\psi_{L}$
\[
\Psi^{L}\star\Fock=\pm\Fock\,.
\]

Let us now consider free field dynamics in the generalized $AdS$
superspace.  The corresponding supersymmetry algebra is
$osp(\L,M)\oplus osp(\L,M)$ while the superspace is $osp(\L,M)$.
To describe background fields (i.e., Cartan forms) in such a space
we follow the same procedure as for $Sp(M)$.

The $OSp(\L|M)$ supergroup is realized by $(M+\L)\times(M+\L)$
matrices $U_{A}{}^{B}$, where $A=(\al,i) (\al=1\dots M,
i=1\dots\L)$, satisfying the group condition
\be
U_{A}{}^{B}U_{C}{}^{D}\Omega^{AC}=\Omega^{BD}\,,
\ee
where $\Omega^{AB}=-(-1)^{\pi_{A}\pi_{B}}\Omega^{BA}$ and
\[
\pi_{A}= \left\{
\begin {array}{ll}
1 \, , \, A=i\\
0 \, , \, A=\al\\
\end {array}
\right. \,.
\]
It can be described by the local supercoordinates
$X^{AB}=(-1)^{\pi_{A}\pi_{B}}X^{BA}$ with the aid of the
exponential parametrization
\be\label{supexp}
U_{A}{}^{B}=\exp(\gl X)_{A}{}^{B}\,.
\ee
Let us introduce the super-oscillators $a_{A}$, $b^{A}$ satisfying
the (anti)commutation relations
\be
a_{A}\star b^{B}-(-1)^{\pi_{A}\pi_{B}}b^{B}\star
a_{A}=\delta_{A}{}^{B}\,,
\ee
with respect to the star-product
\be\label{superstar}
(f\star g)(a,b|X)=\frac{1}{2^{2L}\pi^{M}}\int
f(a+u,b+t)g(a+s,b+v)e^{2(t^{A}s_{A}-v^{A}u_{A})} du dt ds dv\,,
\ee
where the statistics of the integration variables is defined
according to
\[
u_{A}u_{B}=(-1)^{\pi_{A}\pi_{B}}u_{B}u_{A}\,.
\]
The integration measure is chosen so that
 1 is the unit element of the star-product algebra
(\ref{superstar}).

Using the oscillator realization of $osp(\L|M)\oplus
osp(\L|M)\subset osp(2\L|2M)$ we can set
\be
w_{0}=\omega^{AB}a_{B}b_{A}+h^{AB}(a_{B}a_{A}+\gl^{2}b_{B}b_{A})\,.
\ee
The analysis analogous to that of section 2 shows that the gauge
function
\be
g=\sqrt{sdet\|1-\gl^{2}f^{2}(X)\|}\exp\Big
(-f^{AB}(X)(a_{B}a_{A}+\gl^{2}b_{B}b_{A})\Big)
\ee
provides the "Lorentz connection" $\omega_{AB}$ and the "frame"
$h_{AB}$ of the form
\be
\omega_{AB}=\f{1}{2}(dU_{A}{}^{C}(U^{-1})_{CB}+
d(U^{-1})_{A}{}^{C}U_{CB})\,,
\ee
\be
h_{AB}=\f{1}{4\gl}(dU_{A}{}^{C}(U^{-1})_{CB}-
d(U^{-1})_{A}{}^{C}U_{CB})\,,
\ee
where
\be
U_{A}{}^{B}=\Big (\f{1+\gl f(X)}{1-\gl f(X)} \Big )_{A}{}^{B}\,.
\ee

The relationship between $h_{AB}$ and $\omega_{AB}$ is analogous
to (\ref{w(h)})
\be
\omega_{AB}=2h_{A}{}^{C}f_{CB}-2f_{A}{}^{C}
h_{CB}+\gl^{2}f_{A}{}^{C}\omega_{C}{}^{D} f_{DB}\,.
\ee

Here is the list of the gauge functions and corresponding Cartan
forms in different coordinates.

1. The exponential parametrization (\ref{supexp})
\be
g=\f{1}{sdet\|ch\f{\gl X}{2}\|} \exp\Big (-\Big (th\f{\gl
X}{2}\Big )^{AB}(a_{B}a_{A}+\gl^{2}b_{B}b_{A})\Big )\,,
\ee
\be
\omega_{AB}=\f{\gl}{2}\Big (\int\limits_{0}^{1}\exp(\gl X
t)_{A}{}^{C} d X_{C}{}^{D}\exp(\gl X t)_{DB}
dt-\int\limits_{-1}^{0}\exp(\gl X t)_{A}{}^{C} d
X_{C}{}^{D}\exp(\gl X t)_{DB}d t\Big )
\ee
\be
h_{AB}=\f{1}{4}\int\limits_{-1}^{1}\exp(\gl X t)_{A}{}^{C} d
X_{C}{}^{D}\exp(\gl X t)_{DB} dt
\ee

2. $f^{AB}=\phi(sdet\|X\|)X^{AB}$
\be
g=\sqrt{sdet\|1-\gl^{2}\phi^{2}\cdot X^{2}\|}\exp \Big(-\phi
X^{AB}(a_{B}a_{A}+\gl^{2}b_{B}b_{A})\Big)\,,
\ee
\bee
h_{AB}=\phi\cdot\Big (\f{1}{1-\gl^{2}\phi^{2}X^{2}}\Big
)_{A}{}^{C}
(dX_{C}{}^{D}-\gl^{2}\phi^{2}X_{C}{}^{M}dX_{M}{}^{N}X_{N}{}^{D})
\Big(\f{1}{1-\gl^{2}\phi^{2}X^{2}}\Big)_{DB}-\nn\\
-\tilde{\phi}\cdot\Big (\f{X} {1-\gl^{2}\phi^{2}X^{2}}\Big )_{AB}
(X^{-1})_{M}{}^{N}dX_{N}{}^{M}\nn\,,
\eee
\be
\omega_{AB}=2\gl^{2}\phi^{2}\cdot \Big
(\f{1}{1-\gl^{2}\phi^{2}X^{2}}\Big )_{A}{}^{C}
(dX_{C}{}^{M}X_{M}{}^{D}-X_{C}{}^{M}dX_{M}{}^{D})
\Big(\f{1}{1-\gl^{2}\phi^{2}X^{2}}\Big)_{DB}\,,
\ee
where
\be
\tilde{\phi}=\f{\p\phi}{\p\ln sdet\|X\|}\,.
\ee

3. The ``stereographic" coordinates $f^{AB}(X)=\Big
(\f{X}{1\mp\sqrt{1-\gl^{2}X^{2}}}\Big )^{AB}$
\be
g^{\pm}=\sqrt{sdet\|1-\gl^{2}f^{2}(X)\|}\exp\Big (-\Big
(\f{X}{1\mp\sqrt{1-\gl^{2}X^{2}}}\Big )^{AB}
(a_{B}a_{A}+\gl^{2}b_{B}b_{A})\Big)\,,
\ee
\be
h_{AB}=\f{1}{2}\Big (\f{1}{\sqrt{1-\gl^{2}X^{2}}}\Big )_{A}{}^{C}
dX_{C}{}^{D}\Big (\f{1}{\sqrt{1-\gl^{2}X^{2}}}\Big )_{DB}\,.
\ee

In the supersymmetric case, the global higher spin symmetry
transformation law for the generating function $C(b|X)$ with
respect to infinitesimal parameter
\[
\xi=\xi_{0}\exp(\mu^{A}a_{A}-b^{A}\eta_{A})
\]
is analogous to (\ref{transC})
\be
\gd C(b|X)=C(b+\hat{\mu}|X)\exp\Big
(-(b^{A}+\f{1}{2}\hat{\mu}^{A})\hat{\eta}_{A}\Big)\,,
\ee
where
\be
\hat{\mu}^{A}=ch(\gl X)^{AB}\mu_{B}-\f{sh(\gl
X)^{AB}}{\gl}\eta_{B}\,, \quad \hat{\eta}^{A}=ch(\gl
X)^{AB}\eta_{B}-\gl\cdot sh(\gl X)^{AB}\mu_{B}\,.
\ee

It is straightforward to extend the rest of the analysis to the
dynamics in the generalized superspace. Also, having found left
invariant forms it is elementary to write down world-line particle
actions (see e.g., \cite{BLS,tw,osp} for more details and
references). The form of the world-line particle Lagrangian
suggested in \cite{osp} is
\be\label{Lagr}
L=\dot{X}^{AB}w_{0 AB}(a,b|X)+a_{A}\dot{b}^{A}\,,
\ee
where $dX^{AB}w_{0 AB}(a,b|X)=w_{0}(a,b|X)$ is the vacuum 1-form
satisfying the zero-curvature equation (\ref{zero}) and dot
denotes the derivative with respect to the world line parameter.
Applying the Stokes theorem and using (\ref{zero}) the particle
action (\ref{Lagr}) can be rewritten in the string form as an
integral over a two-dimensional surface bounded by a particle
trajectory and parameterized by $\gs^{l}$
\bee
S=\int_{\Sigma^{2}}(w_{0}(a,b|X)\star\wedge
w_{0}(a,b|X)+da_{A}\wedge db^{A}+\nn\\
+(da_{A}\frac{\partial}{\partial a_{A}}+db^{A}\frac{\partial}
{\partial b^{A}})\wedge w_{0}(a,b|X))\,,
\eee
where the pullback is defined as usual
\be
w_{0} (a,b|X)=d\gs^{l}\frac{\partial X^{AB}}{\partial \gs^{l}}
w_{0 AB}(a,b|X),\quad da_{A}=d\gs^{l}\frac{\partial a_{A}}
{\partial \gs^{l}},\quad db^{A}=d\gs^{l}\frac{\partial b^{A}}
{\partial \gs^{l}}\,.
\ee

The problem of calculating Cartan superforms in $osp(1|2M)$
superspace was considered in \cite{tw} where a particular
parametrization was found with bosonic Cartan forms being at most
bilinear in fermionic coordinates. Note that the star-product
algebra formalism simplifies some of the computational problems
being reduced to evaluation of elementary Gaussian integrals.

\section{Conclusions}
It is demonstrated how the  star-product algebra formalism can be
applied to the calculation of the vacuum of fields of the
generalized $AdS$ space associated with $sp(M)\oplus sp(M)$
subalgebra of the recently proposed in \cite{Sp} generalized
conformal symmetry $Sp(2M)$.  The method is universal working
equally well for the supersymmetric case of  $OSp(L,M)$ with
 any $M$ and $L$. The formalism of star-product algebra is shown to
be very efficient  for
 solving free field equations in non-trivial
(generalized conformally flat) geometries in $\M_M$  and
calculating Cartan forms in arbitrary coordinates. Hopefully it
may have applications to formulations of world line
(super)particle dynamics as well as (super)string actions in $M$
theory backgrounds.

\section*{Acknowledgments}
This research was supported in part by INTAS, Grant No.00-1-254,
the RFBR Grant No.02-02-17067 and the RFBR Grant No.00-15-96566.

\appendix
\section{Appendix}
Let us prove that the formula
\be\label{Amain}
g(U)=\frac{2^{\frac{M}{2}}}{\sqrt{\det\left\|U+1\right\|}}
\exp\left(\frac{1}{2}\left(\frac{U-1}{U+1}\right)^{\al\gb}\al_{\al}
\al_{\gb}\right)
\ee
respects the group multiplication law of $Sp(M)$, i.e. that the
formula (\ref{Gom}) solves the equation (\ref{eq}). Let us look
for $U(f)$ in the form
\be
U(f)=\sum_{n=0}^{\infty}a_{n}f^{n}\,,
\ee
where $a_{n}$ are some coefficients. Hence
$U(f_{1})U(f_{2})=\sum\limits_{m,n=0}^{\infty}a_{m}a_{n}f_{1}^{m}f_{2}^{n}$.
Since this expression contains all $f_{1}$ on the left side, and
$f_{2}$ on the right, we have to find such a function $U(f)$ that
$U(f_{1}\circ f_{2})$ contains $f_{1}$ and $f_{2}$ in the correct
order. We have
\be
U(f_{1}\circ f_{2})=
\sum_{m=0}^{\infty}a_{m}\{\sum_{n=0}^{\infty}(-1)^{n}
((f_{2}f_{1})^{n}(1+f_{2})- (f_{1}f_{2})^{n}(1-f_{1}))\}^{m}\,.
\ee
All terms of wrong order must vanish. The analysis of a few first
terms of $U(f_{1}\circ f_{2})$ gives a hint that the coefficients
are: $a_{n}=\{a_{0},a,a,a,...\}$, i.e.
\be
U(f)=a_{0}-a+\frac{a}{1-f}\,.
\ee
The substitution of $U(f)$ into the equation $U^{2}(f)=U(f\circ
f)$ fixes $a_{0}=1, a=2$ so that
\be\label{U(f)}
U(f)=\frac{1+f}{1-f}\,.
\ee
 To prove that the
obtained solution satisfies the equation (\ref{eq}) one has to
check the identity
\be
(1+f_{1}\circ f_{2})\frac{1-f_{2}}{1+f_{2}}=(1-f_{1}\circ
f_{2})\frac{1+f_{1}}{1-f_{1}}\,
\ee
equivalent to the relation
\be\label{ident}
\begin{array}{c}
\{1+\sum\limits_{n=0}^{\infty}(-1)^{n}((f_{2}f_{1})^{n}(1+f_{2})-
(f_{1}f_{2})^{n}(1-f _{1}))\}
(1+2\sum\limits_{m=1}^{\infty}(-1)^{m}f_{2}^{m})=\\
\{1-\sum\limits_{n=0}^{\infty}(-1)^{n}((f_{2}f_{1})^{n}(1+f_{2})-
(f_{1}f_{2})^{n}(1-f _{1}))\}
(1+2\sum\limits_{m=1}^{\infty}f_{1}^{m})\,,
\end{array}
\ee
which is elementary to check.

The normalization factor solves the equation
\be
\frac{r(U_{1})r(U_{2})}{\sqrt{\det\left\|f_{1}f_{2}+1\right\|}}=
r(U_{1}U_{2})\,,
\ee
which is obviously true after the substitution (\ref{Gom})
\be
\frac{2^{\frac{M}{2}}}{\sqrt{\det\left\|U_{1}+1\right\|}}\cdot
\frac{2^{\frac{M}{2}}}{\sqrt{\det\left\|U_{2}+1\right\|}}\cdot
\frac{1}{\sqrt{\det\left\|\frac{U_{1}-1}{U_{1}+1}\frac{U_{2}-1}
{U_{2}+1}+1\right\|}}=
\frac{2^{\frac{M}{2}}}{\sqrt{\det\left\|U_{1}U_{2}+1\right\|}}\,.
\ee
This completes the proof of eq. (\ref{Gom}). The proof for the
supersymmetric case is analogous.


\begin{thebibliography}{99}

\bibitem{osp} M.A. Vasiliev, {\it Phys.Rev.} {\bf D 66}
(2002) 066006, hep-th/0106149.
\bibitem{Sp} M.A.~Vasiliev, Relativity, Causality, Locality,
Quantization and Duality in the $Sp(2M)$ Invariant Generalized
Space-Time, hep-th/0111119.
\bibitem{Fr}C. Fronsdal, Massless Particles, Ortosymplectic
Symmetry and Another Type of Kaluza-Klein Theory, Preprint
UCLA/85/TEP/10, in Essays on Supersymmetry, Reidel, 1986
(Mathematical Physics Studies, v.8).
\bibitem{BLS} I. Bandos, J. Lukierski and  D. Sorokin,
{\it Phys.Rev.} {\bf D61} (2000): 045002,
 hep-th/9904109.
\bibitem{tw} I. Bandos, J. Lukierski, C. Preitschopf, D. Sorokin,
{\it Phys.Rev.} {\bf D61} (2000) 065009, hep-th/9907113.
\bibitem{Bars} I.Bars, C.Deliduman and J.Minic, {\it Phys.Lett.} {\bf
 B466} (1999) 135-143,
 hep-th/9906223.
\bibitem{BG} I.Bars and M.G\"unaydin, {\it Commun.Math.Phys.}
{\bf 91} (1983) 31.
\bibitem{Shayn} O.V. Shaynkman and M.A.~Vasiliev, {\it Theor. Math.
Phys.} {\bf 128} (2001) 1155-1168; ({\it Teor. Mat. Fiz.} {\bf
128} (2001) 378-394), hep-th/0103208.
\bibitem{Bol} K.I. Bolotin and M.A.~Vasiliev, {\it Phys.Lett.} {\bf B479}
(2000) 421-428, hep-th/0001031.
\bibitem{Todor} I. Todorov and G. Mack, {\it Phys.Rev.} {\bf D
6}: 1764-1787, 1973.
\bibitem{Barut} A.O. Barut and R. Raczka Theory of Group
Representations and Applications, {\it PWN-Polish scientific
publishers Warszawa 1977}
\bibitem{Vas} M.A.~Vasiliev, {\it Fortschr.Phys.} {\bf 36} (1988) 33.
\bibitem{Spin} M.A.~Vasiliev, Higher Spin Gauge Theories: Star Product
and $AdS$ Space, hep-th/9910096.
\end{thebibliography}
\end{document}